\documentstyle{article}
\begin{document}
\title{Gravitational waves from coalescing compact
       binaries\thanks{To appear in the
       {\it Proceedings of the Sixth Canadian Conference on
       General Relativity and Relativistic Astrophysics}.}}
\author{Eric Poisson\thanks{Address after September 1, 1995:
        Department of Physics, University of Guelph, Guelph,
        Ontario, Canada N1G 2W1.} \\
        Department of Physics \\
        Washington University \\
        St.~Louis, Missouri 63130, USA}
\date{}
\maketitle
\begin{abstract}
This article is intended to provide a pedagogical account of
issues related to, and recent work on, gravitational waves from
coalescing compact binaries (composed of neutron stars and/or
black holes). These waves are the most promising for kilometer-size
interferometric detectors such as LIGO and VIRGO. Topics discussed
include: interferometric detectors and their noise; coalescing
compact binaries and their gravitational waveforms; the technique
of matched filtering for signal detection and measurement; waveform
calculations in post-Newtonian theory and in the black-hole
perturbation approach; and the accuracy of the post-Newtonian
expansion.

{\it Contents:} 1.~Introduction. 2.~Interferometric detectors.
3.~Detector noise. 4.~More about detector noise. 5.~Coalescing
compact binaries. 6.~Waveform according to the quadrupole
formula. 7.~Matched filtering. 8.~The signal-to-noise ratio.
9.~Signal detection. 10.~Signal measurement. 11.~Waveform
calculations: post-Newtonian theory. 12.~Waveform to second
post-Newtonian order. 13.~Waveform calculations: perturbation
theory. 14.~Luminosity from the perturbation approach.
15.~Accuracy of the post-Newtonian expansion. 16.~Conclusion.
\end{abstract}

\section{Introduction}

The existence of gravitational waves is an unambiguous
prediction of the theory of general relativity
\cite{Thorne87}. Yet,
despite efforts originating in the early nineteen sixties,
gravitational waves have not been detected directly.
Nevertheless, observation of the binary pulsar PSR 1913+16
convinces us that gravitational waves do exist, and that they are
correctly described by Einstein's theory
\cite{Taylor}. That gravitational
waves have not yet been detected on Earth is simply due to their
incredible weakness: typical waves would produce in a
bulk of matter a strain $\Delta L/L$, where $L$ is the
extension of the matter, of order $10^{-21}$
\cite{Thorne87}. Needless to
say, to measure this effect is a great challenge for
experimentalists.

\section{Interferometric detectors}

There is reasonable hope that gravitational waves will be
detected within the next ten years, thanks to a new
generation of detectors which use interferometry to
monitor the small displacements induced by the passage
of a gravitational wave. Two groups are currently involved
in building large-scale interferometers: the American LIGO
team, and the French-Italian VIRGO team.

The LIGO (Laser Interferometer Gravitational-wave Observatory)
project \cite{LIGO} involves two detectors, to be built in Hanford,
Washington, and in Livingston, Louisiana. Construction has
begun at both sites. Each interferometer has an
armlength of approximately 4 km. LIGO should be completed
by the turn of the century.

The VIRGO (so named after the galaxy cluster) project
\cite{VIRGO} involves
a single interferometer, to be built near Pisa, Italy, with
an armlength of approximately 3 km. VIRGO should also be
completed by the turn of the century.

The basic idea behind interferometric detectors is the
following \cite{Thorne87}:

The interferometer is composed of two long
(4 km for LIGO) vacuum pipes forming the letter {\bf L}.
A laser beam is split in two at the corner of the {\bf L},
and is sent into each arm of the interferometer. Each beam
then bounces off a mass which is suspended at each end of the
{\bf L} (a mirror has been coated onto each mass). The light is
finally recombined at the beam splitter, and its intensity is
measured by a photodiode.

When no gravitational wave is present at the interferometer,
the length of each arm is so adjusted that when measured
by the photodiode, the light's intensity is precisely zero
(the recombined beams are arranged to be precisely out
of phase). However, when a gravitational wave passes through
the interferometer, the armlengths are no longer constant,
and the recombined beams no longer precisely out of phase.
More precisely, during the first half of its cycle the
gravitational wave increases the length of one arm, and
decreases the length of the other. During the second half
cycle, the first arm is now shorter, and the second arm
longer. The light's intensity therefore oscillates with
the gravitational-wave frequency.
The intensity is a measure of $\Delta L / L
= h$, where $L$ denotes the interferometer armlength,
and $h$ the gravitational-wave field.

\section{Detector noise}

Interferometers are subject to various sources of noise which
limit the detector's sensitivity to gravitational waves. The
relative importance of each source depends on the frequency at
which the interferometer oscillates \cite{LIGO}.

At low frequencies ($f < 10\ \mbox{Hz}$) the detector's
sensitivity is limited by seismic noise, which is due to
the Earth's seismic activity. At frequencies larger
than 10 Hz the seismic noise can be eliminated with
sophisticated isolation stacks; these fail at low frequencies.

At high frequencies ($f > 100\ \mbox{Hz}$) the detector's
sensitivity is limited by photon shot noise, which is due to
statistical errors in the counting of photons by the photodiode.
This source of noise can be reduced by increasing the laser
power, or making use of ``light recycling'' \cite{LIGO}.

At intermediate frequencies ($f$ between 10 Hz and 100 Hz) the
noise is dominated by thermal noise, which is due to spurious
motions of thermal origin. For example, the suspended masses
are thermally excited and vibrate with their normal-mode
frequencies; this evidently affects the recombined laser beam.

Interferometers are therefore broad-band detectors, with
good sensitivity in the range \cite{LIGO}
\begin{equation}
10\ \mbox{Hz} < f < 1\ 000\ \mbox{Hz}.
\label{1}
\end{equation}
The required sensitivity for full-scale interferometers
is approximately $h_n \sim 10^{-22}$ at peak sensitivity ---
a tall order. [The subscript $n$ stands for ``noise level'';
we will define $h_n(f)$ precisely below. A plot of $h_n(f)$,
appropriate for an interferometric detector with ``advanced''
sensitivity, is given in Fig.~1.] For comparison,
we may mention that the Caltech 40 m prototype has already
achieved $h_n \simeq 10^{-19}$ at peak sensitivity
$(f=450\ \mbox{Hz})$. It is not implausible that
improved technology and a factor of 100 in armlength
will permit to reach the desired goal.

\section{More about detector noise}

\begin{figure}[t]
\special{hscale=50 vscale=50 hoffset=60.0 voffset=-300.0
         angle=0.0 psfile=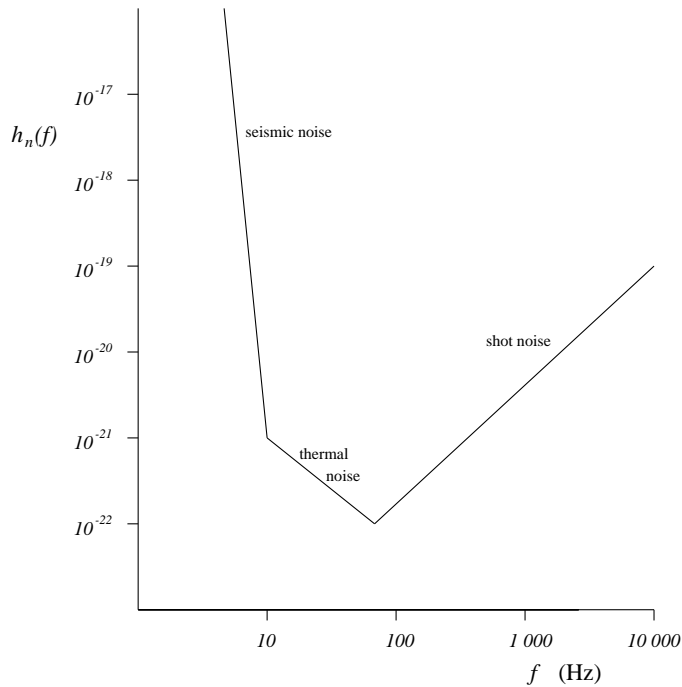}
\vspace*{3.2in}
\caption[Fig. 1]{Noise level in an interferometric
detector with advanced sensitivity.}
\end{figure}

The detector noise can be measured when no gravitational
wave is present at the interferometer, the typical
situation. Then the detector
output $s(t) = \Delta L(t)/L$ is given by noise alone:
\begin{equation}
s(t) = n(t),
\label{2}
\end{equation}
where $n(t)$ represents the noise. The noise is a
random process
\cite{Reif}: the function $n(t)$ takes purely random
values. Consequently, the noise can only be studied
using statistical methods.
In the presence of a gravitational wave, Eq.~(\ref{2}) must
be replaced by $s(t) = h(t) + n(t)$, where $h(t)$ is the
gravitational-wave field.

The statistical properties of the noise can be
determined by careful measurement. For example,
the time average
\begin{equation}
\overline{n(t)} =
\lim_{T\to\infty}\, \frac{1}{2T} \int_{-T}^{+T}
n(t)\, dt
\label{3}
\end{equation}
can be constructed. This mean value can then be subtracted
from $n(t)$ and, without loss of generality, we can
put $\overline{n} = 0$. Also from measurements, the
noise's {\it autocorrelation function} $C_n(\tau)$
can be constructed:
\begin{equation}
C_n(\tau) = \overline{n(t) n(t+\tau)};
\label{4}
\end{equation}
$C_n(0)$ gives the mean squared deviation of the
noise with respect to the mean value.

In the following we will assume that the noise is
{\it stationary}, in the sense that its statistical
properties do not depend on time \cite{Reif}. This means, in
particular, that the autocorrelation function does
not depend explicitly on the origin of time $t$, but only on
the variable $\tau$, as was expressed in Eq.~(\ref{4}).
We shall also assume that the noise satisfies the
{\it ergodic hypothesis}, so that time averages can
be replaced with ensemble averages \cite{Reif}. Here, the noise
is imagined to be drawn from a representative ensemble,
and the probability that it takes a particular realization
$n(t)$ is given by a specified probability distribution.
The statistical properties of the noise then refer to this
(infinite dimensional) distribution function.

In full generality, the statistical properties of the
noise can only be summarized by constructing all the
higher moments $\overline{n \cdots n}$. If, however,
the noise is assumed to be Gaussian, in the sense that
its probability distribution function is an infinite
dimensional Gaussian distribution \cite{Helstrom},
then the autocorrelation function contains all the
information.

Real detector noise is neither strictly stationary nor
strictly Gaussian. However, on a timescale of hours,
which is long compared with typical gravitational-wave
bursts, the noise appears stationary to a good approximation
\cite{LIGO}.
And non-Gaussian components to the noise can be removed,
to a large extent, by cross-correlating detector outputs
from two widely separated interferometers
\cite{LIGO}. It is therefore
a satisfactory approximation to take the noise to be
stationary and Gaussian.

Under these assumptions the statistical properties of the
detector noise are fully summarized by $C_n(\tau)$.
It is convenient to work instead in
the frequency domain, and to define \cite{Reif} the noise's
{\it spectral density} $S_n(f)$ as
\begin{equation}
S_n(f) = 2 \int C_n(\tau) e^{2\pi i f \tau}\, d\tau.
\label{5}
\end{equation}
The spectral density is defined for $f>0$ only;
as $C_n(\tau)$ is a real and even function, the
negative frequencies only duplicate the information
contained in the positive frequencies.

As $n(t)$ is dimensionless, the spectral density has
dimensions of time. By multiplying $S_n(f)$ with the
frequency and taking the square root (since the
spectral density represents the mean squared
noise), one obtains the {\it noise level} $h_n(f)$:
\begin{equation}
h_n(f) = \sqrt{ f S_n(f) }.
\label{6}
\end{equation}
This gives the equivalent gravitational-wave amplitude
which would make the interferometer oscillate at just
the noise level \cite{Thorne87};
this quantity was introduced in the
preceding section (see Fig.~1).

\section{Coalescing compact binaries}

Coalescing compact binaries, composed of neutron stars
and/or black holes, are the most promising source of
gravitational waves for interferometric detectors
\cite{Thorne87,Schutz}.

Consider the binary pulsar PSR 1913+16
\cite{Taylor}. This system
consists of two neutron stars, each of $1.4\ M_\odot$,
in orbital motion around each other. Its present
orbital period $P$ is approximately 8 hours, corresponding
to orbital separations of about $5\times 10^5$ times
the total mass. (Here and throughout we use units such that
$G=c=1$.) Its present eccentricity is approximately
equal to 0.6.

The binary's orbital period is observed
to decay at a rate $dP/dt = -2 \times 10^{-12}$
corresponding precisely to a loss of energy and angular
momentum to gravitational waves \cite{Taylor}.
In a timescale of approximately $10^8$ years the orbital
period will have decreased to less than a tenth of a second,
corresponding to orbital separations smaller than
one hundred times the total mass. In this time, the
eccentricity will have been reduced to extremely small
values (by the radiation reaction), so that the
orbits are practically circular. The gravitational
waves produced then have a frequency larger than 10 Hz, and
the frequency keeps increasing as the system evolves.
{\it During this late stage of orbital evolution, the
gravitational waves sweep through the frequency bandwith
of interferometric detectors, and thus become visible.}
By the time the orbital separation becomes as small as
a few times the total mass, the neutron stars begin to
merge. The gravitational waves produced during the final
merger cannot be detected by interferometric detectors,
at least in the broad-band configuration described
above \cite{foot1}:
the gravitational-wave frequency is then larger than 1 000 Hz,
for which the detector noise is large.

Of course, it would be foolish to wait $10^8$ years in order
for PSR 1913+16 to produce gravitational waves with appropriate
frequency. Fortunately, interferometric detectors will be
sensitive enough to monitor binary coalescences
occurring in quite a large volume of the universe,
approximately $10^7\ \mbox{Mpc}^3$ (corresponding to
a radius of 200 Mpc \cite{LIGO}).
It has been estimated \cite{Phinney} that as many
as 100 coalescences could occur every year in such a volume
(this includes coalescences of black-hole
systems as well). This potentially large event rate is one
of the factors that make gravitational waves from
coalescing compact binaries especially promising.

The other factor comes from the fact that compact
binaries are extremely clean astrophysical systems.
It can be estimated \cite{Kochanek} that tidal interactions
between the two stars are completely negligible, up
to the point where the objects are about to merge.
In particular, no mass transfer occurs. The system
can therefore be modeled, to extremely good accuracy,
as that of two point masses with a limited number of
internal properties (such as mass, spin, and quadrupole
moment). The challenge in modeling coalescing compact binaries
resides in formulating and solving the equations of
motion and wave generation for a general relativistic
two-body problem \cite{Damour}.

\section{Waveform according to the quadrupole formula}

At the crudest level, the gravitational-wave signal
corresponding to an inspiraling binary system can be
calculated by (i) assuming that the orbital motion is
Newtonian (with the effects of radiation reaction
incorporated) and (ii) using the standard quadrupole
formula for wave generation \cite{MTW}.
As motivated above, we
may also assume that the orbits are circular.

We define $h(t)$ to be the gravitational-wave signal.
This is given by a linear combination, appropriate
for interferometric detectors, of the two
fundamental polarizations, $h_+$ and $h_\times$,
of the gravitational-wave field. At this
level of approximation, the signal is given
by \cite{Thorne87}
\begin{equation}
h(t) = Q(\mbox{angles})\, ({\cal M}/r)\,
(\pi {\cal M} f)^{2/3}\, \cos \Phi(t).
\label{7}
\end{equation}
As expected, the signal decays as the inverse power
of $r$, the distance to the source.

In Eq.~(\ref{7}), $Q$ is a function of all the angles
relevant to the problem: position of the source in the
sky, orientation of the orbital plane, position and
orientation of the detector on Earth. The parameter
${\cal M}$ is called the {\it chirp mass} and
represents a particular combination of the masses,
given by
\begin{equation}
{\cal M} = (m_1 m_2)^{3/5} / (m_1 + m_2)^{1/5}.
\label{8}
\end{equation}
The waveform depends on the chirp mass only, and not
on any other combination of the masses. The symbol
$f$ represents the gravitational-wave frequency,
which is equal to {\it twice} the orbital frequency.
Because the system loses energy and angular momentum
to gravitational waves, the frequency is not constant,
but increases in time according to \cite{Thorne87}
\begin{equation}
\frac{df}{dt} = \frac{96}{5\pi {\cal M}^2}
(\pi {\cal M} f)^{11/3}.
\label{9}
\end{equation}
As a consequence, Eq.~(\ref{7}) shows that the
amplitude of the signal, which is proportional
to $(\pi {\cal M} f)^{2/3}$, also increases with
time. A signal which increases both in frequency
and in amplitude is known as a {\it chirp}, and this is
the origin of the term ``chirp mass''. Finally,
the phase function $\Phi(t)$ is given by
\begin{equation}
\Phi(t) = \int^t 2\pi f(t')\, dt'.
\label{10}
\end{equation}
Because the frequency is not a constant, the
phase accumulates nonlinearly with time.

For a system of two neutron stars, the gravitational-wave
signal undergoes approximately 16 000 oscillations
as it sweeps through the frequency bandwith of an
interferometric detector. The timescale for the
frequency sweep is approximately 15 minutes.
The orbital separation ranges from approximately
180 to 10 times the total mass
$M = m_1 + m_2$, and the orbital velocity
\begin{equation}
v \equiv (\pi M f)^{1/3}
\label{11}
\end{equation}
ranges from approximately 0.1 to 0.4. This indicates that
relativistic corrections must be inserted in
Eqs.~(\ref{7}) and (\ref{9}) in order
to obtain a satisfactory degree of accuracy.

\section{Matched filtering}

How does one go about finding a gravitational-wave signal
in a noisy data stream, when typically the signal is not
very strong? And once the signal is found, how does one
go about extracting the information it contains? For
signals of precisely known form, one goes about this using
the technique of {\it matched filtering} \cite{Wainstein}.
Signals from inspiraling compact binaries, since they
can be calculated with high precision, belong to this class.
The basic idea behind matched filtering is to use our
knowledge about the signal in order to go find it in the
data stream, after the noisy frequencies have been filtered
out.

Suppose a signal of known form $h(t;\vec{\mu})$ is
present in the data stream. Here, the vector
$\vec{\mu}$ collectively denotes all the parameters
characterizing the signal. In the case of inspiraling
binaries, these would be the time of arrival, the initial
phase, the distance to the source, the position angles,
the chirp mass, and other parameters to be introduced
in Sec.~12. The detector output is given by
\begin{equation}
s(t) = h(t;\vec{\mu}) + n(t),
\label{12}
\end{equation}
where $n(t)$ is the noise, whose statistical properties
are fully summarized by the spectral density $S_n(f)$, as
discussed in Sec.~4.

The first step in matched filtering
\cite{CutlerFlanagan} is to pass $s(t)$
through a linear filter which removes the noisy
frequencies. The idea here is to use our knowledge
about the detector noise contained in $S_n(f)$ in
order to discard that part of the detector output for
which the detector noise is large. The output is imagined
to be decomposed into Fourier modes according to
$s(t) = \int \tilde{s}(f) e^{-2\pi i f t}\, df$;
the filter suppresses the modes $\tilde{s}(f)$
for which $S_n(f)$ is large.

In mathematical terminology, a linear filter is a linear
operation on a function $s(t)$. This operation can
always be written as \cite{Wainstein}
\begin{equation}
s(t) \to \int w(t-t') s(t')\ dt',
\label{13}
\end{equation}
where $w(t-t')$ is the filter function. The filter
which removes the noisy frequencies is the one
such that
\begin{equation}
\tilde{w}(f) = \frac{2}{S_n(|f|)};
\label{14}
\end{equation}
the factor of 2 is conventional.

The next step in matched filtering consists of computing
the overlap integral between the filtered output (\ref{13})
and the known signal $h(t;\vec{\mu}_{\rm trial})$. The
true value $\vec{\mu}$ of the source parameters is not
known prior to the measurement. This operation must therefore
be repeated for a large number of trial values
$\vec{\mu}_{\rm trial}$; the corresponding signals
$h(t;\vec{\mu}_{\rm trial})$ are known as {\it
templates}. The overlap integral defines
the function $S(\vec{\mu}_{\rm trial})$ given by
\begin{eqnarray}
S(\vec{\mu}_{\rm trial}) &=& \int
h(t;\vec{\mu}_{\rm trial}) w(t-t') s(t')\, dt'\, dt
\nonumber \\
&=& \bigl\langle h(\vec{\mu}_{\rm trial}) \bigm| s
\bigr\rangle.
\label{15}
\end{eqnarray}
To obtain the second line we have inserted the Fourier
decompositions of $h$, $w$, and $s$, and carried out
the integrations over time. The inner product
$\langle \cdot | \cdot \rangle$ is defined as
\begin{equation}
\langle a | b \rangle = 2 \int_0^\infty
\frac{\tilde{a}^*(f) \tilde{b}(f) +
      \tilde{a}(f) \tilde{b}^*(f)}{S_n(f)}\, df,
\label{16}
\end{equation}
where $a$ and $b$ are arbitrary functions of time,
with Fourier transforms $\tilde{a}$ and $\tilde{b}$.

A quantity analogous to $S(\vec{\mu}_{\rm trial})$
can be defined for the
noise alone, by replacing $s$ to the right of
Eq.~(\ref{15}) by $n$. Operationally, this amounts to
filtering the detector output when a gravitational-wave
signal is known {\it not} to be present. Because the
noise is a random function, the integrals
$\int h(t;\vec{\mu}_{\rm trial})
w(t-t') n(t')\, dt'\, dt$ are random also.
And because the noise is assumed to have zero mean, to
take an average over all possible realizations of the
noise (by repeating the measurements many times) would
yield a zero value. None of these
quantities would be especially useful. We are therefore
led to consider the root-mean-square average of these
integrals,
\begin{eqnarray}
N(\vec{\mu}_{\rm trial}) &=& \mbox{rms} \int
h(t;\vec{\mu}_{\rm trial}) w(t-t') n(t')\, dt'\, dt
\\ \nonumber
&=& \sqrt{ \bigl\langle h(\vec{\mu}_{\rm trial}) \bigm|
h(\vec{\mu}_{\rm trial}) \bigr\rangle } \equiv
\rho(\vec{\mu}_{\rm trial}).
\label{17}
\end{eqnarray}
To go from the first to the second line requires some
machinery which will not be presented in this review.
We refer the reader to Ref.~\cite{Wainstein} for
the missing steps.

\section{The signal-to-noise ratio}

The {\it signal-to-noise ratio} is defined to be the
ratio of $S$ over $N$:
\begin{equation}
\mbox{\small SNR}(\vec{\mu}_{\rm trial}) = \frac{
\bigl\langle h(\vec{\mu}_{\rm trial})
\bigm| s \bigr\rangle}{\rho(\vec{\mu}_{\rm trial})}.
\label{18}
\end{equation}
It is clear that $\mbox{\small SNR}(\vec{\mu}_{\rm trial})$
is a random function, since $s(t)$ is itself a random function,
being the superposition of signal plus noise.
Its expectation value, or average over all possible
realizations of the noise, is zero in the absence of
signal (since $\overline{n} = 0$), and
\begin{equation}
\overline{\mbox{\small SNR} (\vec{\mu}_{\rm trial})}
= \frac{ \bigl\langle h(\vec{\mu}_{\rm trial})
\bigm| h(\vec{\mu}) \bigr\rangle}{
\rho(\vec{\mu}_{\rm trial})}
\label{19}
\end{equation}
in the presence of the signal $h(t;\vec{\mu})$. It
is important to notice that in Eq.~(\ref{19}), the
numerator is the overlap integral between the
true signal $h(t;\vec{\mu})$ and the templates
$h(t;\vec{\mu}_{\rm trial})$. In the absence
or presence of a signal, the variance in the
signal-to-noise ratio is precisely equal to
unity \cite{CutlerFlanagan},
independently of the values of $\vec{\mu}$ and
$\vec{\mu}_{\rm trial}$. This fully summarizes the
statistical properties of the signal-to-noise ratio,
since $\mbox{\small SNR}(\vec{\mu}_{\rm trial})$
is a Gaussian random function. [This can be seen
from the fact that $s(t)$ itself is Gaussian,
being the superposition of signal plus Gaussian
noise.]

It is intuitively clear that choosing a template
with $\vec{\mu}_{\rm trial} = \vec{\mu}$
will produce the largest possible expectation
value of the signal-to-noise ratio. This statement,
known as {\it Wiener's theorem}
\cite{Wainstein}, can easily be
shown to be true by applying Schwarz's inequality
to the right-hand side of Eq.~(\ref{19}). We have
\begin{equation}
\mbox{max} \Bigl\{ \overline{ \mbox{\small SNR}
(\vec{\mu}_{\rm trial})} \Bigr\} =
\overline{\mbox{\small SNR} (\vec{\mu})} = \rho(\vec{\mu}) =
\sqrt{ \bigl\langle h(\vec{\mu}) \bigm|
h(\vec{\mu}) \bigr\rangle }.
\label{20}
\end{equation}
The fact that the signal-to-noise ratio has a variance of unity
indicates that a signal can be concluded to be present only
if $\rho(\vec{\mu})$ is significantly larger than 1. We will
come back to this point in the next section.

The true value of the source parameters can therefore
be determined by maximizing $\mbox{\small SNR}
(\vec{\mu}_{\rm trial})$ over all possible values of
the trial parameters. The fact that the signal-to-noise
ratio has a variance of unity implies that this
determination can only have
a limited degree of accuracy. The statistical
errors decrease with increasing $\rho(\vec{\mu})$; since
this is proportional to the signal's amplitude,
a stronger signal gives better accuracy.

Maximizing the signal-to-noise ratio is essentially
equivalent to maximizing the overlap integral
\[
\bigl\langle h(\vec{\mu}_{\rm trial})
\bigm| h(\vec{\mu}) \bigr\rangle.
\]
It is easy to see how the choice of the parameters
$\vec{\mu}_{\rm trial}$ can affect the overlap
integral. Consider a toy waveform with three parameters:
arrival time, initial phase, and chirp mass. A mismatch in
the arrival times clearly reduces the overlap integral: the
signal and the template, taken to be functions of time, might
have support in entirely different regions of the time axis,
leading to a vanishing overlap. Supposing that the arrival
times are matched, a mismatch in the initial phases can also
reduce the overlap integral: the signal and the template
might be out of phase with each other, leading
to an oscillating integrand and a vanishingly small overlap.
Finally, supposing that both the arrival times and initial
phases are matched, a mismatch in the chirp masses would also
reduce the overlap integral. This is because the chirp mass
governs the rate at which the signal's frequency changes with
time; cf.~Eq.~(\ref{9}). Signal and template, starting
at the same time with the same phase, might thereafter
go out of phase, thereby reducing the overlap.

\section{Signal detection}

The first order of business when analysing the output
of a gravitational-wave detector is to
decide whether or not a signal is present. Here
we assume that the signal must be of a specific
form, corresponding to a coalescing binary
system. In this section we discuss signal detection
--- how the technique of matched filtering can be
employed to find the signal in the noisy data stream.
In the next section we will discuss signal
measurement --- how matched filtering is used to estimate
the value of the source parameters once the signal has been
found.

As mentioned in the previous section, a signal can be concluded
to be present if the maximum
value of the signal-to-noise ratio, $\mbox{\small SNR} \equiv
\mbox{max}\{\mbox{\small SNR}(\vec{\mu}_{\rm trial})\}$, is
significantly larger than unity. In fact, there
exists a threshold value $\mbox{\small SNR}^*$
such that a signal is concluded to present,
with a certain confidence level,
if $\mbox{\small SNR} > \mbox{\small SNR}^*$. To
figure out how large this threshold must be is
a standard application of the statistical theory
of signal detection \cite{Helstrom}, which was developed largely
for the purpose of detecting radar signals. The
theory can easily be taken over to the case gravitational-wave
signals \cite{Finn}. To go into the detail of this theory
would be outside the scope of this review.
We shall simply state that $\mbox{\small SNR}^*$ is
fixed by selecting a small, acceptable value for the
probability that a signal would falsely be concluded
to be present --- the false alarm probability. (This is the
Neyman-Pearson criterion \cite{Helstrom}, which is more
precisely formulated in terms of the likelihood ratio,
the ratio of the probability that a signal is present
to the probability that it is absent.)
Once $\mbox{\small SNR}^*$ is
fixed, the level of confidence that a signal is
indeed present increases with
$\mbox{\small SNR} > \mbox{\small SNR}^*$. A
typical ballpark value for the threshold is
$\mbox{\small SNR}^* = 6$.

In the preceding paragraph it was assumed that the
signal-to-noise ratio
$\mbox{\small SNR}$ is computed using template
waveforms which are functions of the parameters
$\vec{\mu}$ introduced in Sec.~7.
These parameters have direct physical meaning; they
include the chirp mass and other meaningful parameters
to be introduced in Sec.~12. However, since these
parameters are {\it not} estimated during the detection
stage of the data analysis (they are estimated only
{\it after} a signal has been found), there is no
particular need to parametrize the templates with $\vec{\mu}$.
In fact, it may be desirable,
in order to minimize the computational
effort, to parametrize the signal in a completely
different way. The new parameters, $\vec{\alpha}$,
would then have no particular physical significance.
What is required is that the new templates,
$h(t;\vec{\alpha})$, reproduce the behaviour of
the expected gravitational-wave signal. In other
words, the templates $h(t;\vec{\mu})$ and
$h(t;\vec{\alpha})$ should span the same
``signal space'', but $h(t;\vec{\alpha})$
should do so most efficiently. Unlike
$h(t;\vec{\mu})$, $h(t;\vec{\alpha})$ need
not be derived from the field equations of
general relativity. With these new
{\it detection templates}, the signal-to-noise
ratio is defined as
\begin{equation}
\mbox{\small SNR} =
\mbox{max} \bigl\{ \mbox{\small SNR}
(\vec{\alpha}) \bigr\},
\label{21}
\end{equation}
where $\mbox{\small SNR}(\vec{\alpha})$ is defined as in
Eq.~(\ref{18}). As before, a signal is concluded
to be present if $\mbox{\small SNR}
> \mbox{\small SNR}^*$.

\section{Signal measurement}

Once the detection templates $h(t;\vec{\alpha})$
have been used to conclude that a signal
is present in the data stream, they are
replaced with the {\it measurement templates}
$h(t;\vec{\mu}_{\rm trial})$ in order to
estimate the value of the physical parameters
$\vec{\mu}$.

The procedure to estimate the source parameters
was explained in Sec.~8.
As was mentioned, the idea
is to maximize over all possible values of the trial
parameters the overlap integral
\[
\bigl\langle h(\vec{\mu}) \bigm|
h(\vec{\mu}_{\rm trial}) \bigr\rangle .
\]
We already have
discussed the effect of a mismatch in the value of
the parameters. What remains to be discussed is
the effect of a mismatch in the functional form
of the template with respect to that of the true
waveform.

The true waveform is governed by the exact laws of
general relativity. The template, on the
other hand, is necessarily constructed using an
approximation to the exact laws. (That approximations
must be made in analytic calculations is obvious; in
a numerical treatment the approximation resides in
the finite differencing of the field equations.)
This, clearly, must have an effect on our
strategy for extracting the information contained
in the gravitational-wave signal. This can be
seen simply from the fact that Wiener's theorem,
as stated in Sec.~8, strictly requires signal
and template to have the same functional form;
they are allowed to differ only in the value of
their parameters.

Suppose that gravitational waves coming from a
given source are received without noise, so that
the true signal $h(t;\vec{\mu})$ is measured accurately.
Suppose also that the true value $\vec{\mu}$ of the
source parameters is known (God has spoken). Then
a computation of the overlap integral with
$\vec{\mu}_{\rm trial} = \vec{\mu}$ but with
an approximate template $h(t;\vec{\mu}_{\rm trial})$
will not yield the maximum possible value for
$\mbox{\small SNR}(\vec{\mu}_{\rm trial})$.
This is because the approximation differs from the
true signal, both in amplitude and in phase. Since both
signal and template undergo a large number of oscillations
(recall that for a system of two neutron
stars, this number is approximately 16 000), the overlap
integral is most sensitive to phase differences: a
slight phase lag causes the integrand to oscillate,
thereby severely reducing the signal-to-noise ratio
with respect to its maximum possible value.

A gravitational-wave astronomer doesn't know before the
measurement the true value of the source parameters, and must
work with an approximation to the true general-relativistic
waveform. We have seen that the phase lag occurring between
the true signal and the approximate template when the parameters
are matched reduces the signal-to-noise ratio from its
maximum possible value. It follows that maximizing
$\mbox{\small SNR}(\vec{\mu}_{\rm trial})$
with approximate templates introduce
{\it systematic errors} into the estimation of
the source parameters: evaluating the signal-to-noise
ratio with $\vec{\mu}_{\rm trial} = \vec{\mu}
+ \delta \vec{\mu}$ will return,
for some $\delta \vec{\mu}$, a number larger
than $\mbox{\small SNR}(\vec{\mu})$. The systematic errors are
precisely the value of $\delta \vec{\mu}$ for which
the signal-to-noise ratio is largest. If the templates are
a poor approximation to the true signal, then
the systematic errors will be larger than the statistical
errors arising because
$\mbox{\small SNR}(\vec{\mu}_{\rm trial})$
is a random function (see Sec.~8).

We therefore appreciate the need for constructing measurement
templates which are as accurate as possible, especially
in phase \cite{FinnChernoff,Cutleretal}.
The requirement is that the systematic errors
in the estimated parameters must be smaller than the
statistical errors. An estimate for the required degree of
accuracy comes from the observation that the overlap
integral will be significantly reduced if the template
loses phase by as much as one wave cycle with respect to
the true signal. Since the total number of wave cycles is
approximately 16 000, we have
\begin{equation}
\mbox{accuracy} \sim \frac{1}{16\ 000} \sim
10^{-4}.
\label{22}
\end{equation}
Since the orbital velocity $v$ is of order $10^{-1}$
when the gravitational-wave frequency is in the relevant
bandwidth, relativistic corrections {\it at
least} of order $v^4$ are required to improve the
quadrupole-formula expression given in Sec.~6.
As we shall see, this is an underestimate.

\section{Waveform calculations: post-Newtonian theory}

We have seen that accurate measurement templates are
required to make the most of the gravitational-wave
signals originating from coalescing compact binaries.
We also have seen that the quadrupole-formula waveform,
Eq.~(\ref{7}), is not sufficiently accurate; relativistic
corrections at least of order $v^4$ are required. How
does one go about calculating these?

A possible line of approach is to use a slow-motion
approximation to the equations of general relativity.
This is based on the requirement that if $v$ is a
typical velocity inside the matter source, then
\begin{equation}
v \ll 1.
\label{23}
\end{equation}
We shall call this approximation ``post-Newtonian
theory'' \cite{Will}. We point out that for binary systems,
post-Newtonian theory makes no assumption regarding
the relative size of the two masses. This is to
be contrasted with the perturbation approach,
discussed in Sec.~13, in which the mass ratio
is assumed to be small, but no restriction
is put on $v$.

In post-Newtonian theory \cite{BlanchetDamour}, one starts
by defining fields $h^{\alpha\beta}$ as
\begin{equation}
h^{\alpha\beta} = \sqrt{-g}\, g^{\alpha\beta}
- \eta^{\alpha\beta},
\label{24}
\end{equation}
where $g^{\alpha\beta}$ is the inverse of the
true metric $g_{\alpha\beta}$ with determinant $g$, and
$\eta^{\alpha\beta}$ is the metric of Minkowski spacetime.
When the harmonic gauge conditions
\begin{equation}
\partial_\beta h^{\alpha\beta} = 0
\label{25}
\end{equation}
are imposed, the {\it exact} Einstein field equations
reduce to
\begin{equation}
\Box h^{\alpha\beta} = 16\pi (-g) T^{\alpha\beta}
+ \Lambda^{\alpha\beta}.
\label{26}
\end{equation}
Here, $\Box=\eta^{\alpha\beta} \partial_\alpha
\partial_\beta$ is the flat-spacetime wave operator,
$T^{\alpha\beta}$ is the stress-energy
tensor of the source, and $\Lambda^{\alpha\beta}$
is nonlinear in $h^{\alpha\beta}$
and represents an effective stress-energy
tensor for the gravitational field.

The detailed way in which one solves these equations is
quite complicated, and will not be described here. The
essential ideas are these \cite{BlanchetDamour}:

One first integrates the equations in the near zone
($r < \lambda$, where $r$ is the flat-space radius
and $\lambda$ the gravitational wavelength) assuming
slow motion, or $\partial h^{\alpha\beta} / \partial t
\ll \partial h^{\alpha\beta} / \partial x^i$. One does
this by iterations: the nonlinear terms in
Eq.~(\ref{26}) are first neglected, and the
resulting linear equations integrated. These solutions
are then used as input for the next iteration. This
process is continued until the desired degree of
accuracy is obtained. This is the standard post-Newtonian
approach \cite{Damour}.

One next integrates the equations everywhere in the
vacuum region outside the source. This is done once
again by iterations, assuming $h^{\alpha\beta} \ll 1$,
but assuming nothing about the relative size of
$\partial h^{\alpha\beta} / \partial t$ with
respect to the spatial derivatives. This is because
the vacuum region contains the wave zone, in which
the field propagates with the speed of light; a
slow-motion assumption would therefore not do for
the field itself. This is the
post-Minkowskian approach \cite{Damour}.

Using the post-Minkowskian approach one constructs,
by successive approximations, the
most general solution to the Einstein equations outside
the source. This is characterized by two infinite
sets of arbitrary multipole moments
\cite{Thorne80}, the mass moments
$M_{\ell m}(t-r)$ and the current moments $J_{\ell m}(t-r)$,
were $\ell$ and $m$ are the standard spherical-harmonic indices.
(In practice, the fields $h^{\alpha\beta}$ are equivalently
expressed in terms of symmetric-trace-free
moments, not spherical-harmonic
moments.) The general solution is then matched to the
near-zone solution in the region of common validity, and the
multipole moments are thus determined.

Finally, one expresses the radiation field ---
the time-varying, $O(1/r)$ part of the gravitational
field --- in terms of the derivatives of the mass and current
multipole moments \cite{Thorne80}.
This gives the gravitational waveform.
The gravitational-wave luminosity $dE/dt$ can also be obtained
from the radiation field.

\section{Waveform to second post-Newtonian order}

To date, the post-Newtonian calculation of the waveform
has been carried out accurately through
second post-Newtonian order --- $O(v^4)$ --- beyond
the leading-order, quadrupole-formula expressions
given in Sec.~6. The complete waveform will not
be displayed here. Instead, we will focus solely
on the waveform's {\it phasing}.

The phasing of the waves can be determined from
$df/dt$, the rate of change of the gravitational-wave
frequency. This can be expressed as
\begin{equation}
\frac{df}{dt} = \frac{dE/dt}{dE/df},
\label{27}
\end{equation}
where $dE/dt$ is (minus) the gravitational-wave luminosity,
and $dE/df$ relates orbital energy to orbital frequency
($f$ is twice the orbital frequency; the orbits are assumed
to be circular). Both these quantities
can be expanded in powers of
\begin{equation}
v \equiv (\pi M f)^{1/3},
\label{28}
\end{equation}
with leading-order terms \cite{MTW}
\begin{equation}
\biggl( \frac{dE}{dt} \biggr)_{\!\!N} = -\frac{32}{5}
\eta^2 v^{10}, \qquad
\biggl( \frac{dE}{df} \biggr)_{\!\!N} = -\frac{\pi}{3}
\mu M v^{-1},
\label{29}
\end{equation}
where the subscript $N$ stands for ``Newtonian''. We
have introduced the reduced mass $\mu$, the total mass
$M$, and the mass ratio $\eta$ as
\begin{equation}
\mu = \frac{m_1 m_2}{m_1 + m_2}, \qquad
M = m_1 + m_2, \qquad
\eta = \frac{\mu}{M}.
\label{30}
\end{equation}
Notice that $\eta$ is restricted to the interval
$0 < \eta \leq 1/4$, with $\eta=1/4$ for $m_1=m_2$.
It is easy to check that Eqs.~(\ref{27}) and (\ref{29})
reproduce Eq.~(\ref{9}) above; the chirp mass can be
expressed as ${\cal M} = \eta^{3/5} M$.

For simplicity we will, in the following, focus on the
quantity $dE/dt$. To second post-Newtonian order,
the luminosity takes the form
\begin{equation}
\frac{dE}{dt} = \biggl( \frac{dE}{dt} \biggr)_{\!\!N}
\Biggl[ 1 -
\biggl( \frac{1247}{336} + \frac{35}{12}\, \eta \biggr) v^2 +
\bigl( 4\pi - \mbox{\small SO} \bigr) v^3 - \biggl(
\frac{44711}{9072} - \frac{9271}{504}\, \eta -
\frac{65}{18}\, \eta^2 + \mbox{\small SS} \biggr) v^4 + \cdots
\Biggr].
\label{31}
\end{equation}
Here, the terms {\small SO} and {\small SS} are due to spin-orbit
and spin-spin interactions, respectively \cite{KWW}; these occur if
the masses $m_1$ and $m_2$ are rotating. Let $\vec{S}_1$
and $\vec{S}_2$ be the spin angular momentum of each mass,
and define the dimensionless quantities $\vec{\chi}_a =
\vec{S}_a/{m_a}^2$, for $a=\{1,2\}$. Let also $\hat{L}$
be the direction of orbital angular momentum. Then \cite{KWW}
\begin{equation}
\mbox{\small SO} = \frac{1}{4} \sum_{a=1}^2 \Bigl[
11 (m_a/M)^2 + 5\eta \Bigr] \hat{L} \cdot \vec{\chi}_a
\label{32}
\end{equation}
is the spin-orbit term, and
\begin{equation}
\mbox{\small SS} = \frac{\eta}{48} \Bigl(
103 \vec{\chi_1} \cdot \vec{\chi_2} -
289 \hat{L} \cdot \vec{\chi}_1 \,
\hat{L} \cdot \vec{\chi}_2 \Bigr)
\label{33}
\end{equation}
is the spin-spin term.

In Eq.~(\ref{31}), the leading-order term was first calculated
in 1963 by Peters and Mathews
\cite{PetersMathews} using the standard quadrupole
formula. The first post-Newtonian correction, at order $v^2$,
was calculated in 1976 by Wagoner and Will
\cite{WagonerWill}. The $4\pi v^3$
term is due to wave propagation effects: As the waves
propagate out of the near zone they are scattered by
the curvature of spacetime, and this modifies both the
amplitude and the phase of the waveform; this ``tail term''
was first calculated in 1993 by this author
\cite{paperI}, and then
independently by Wiseman
\cite{Wiseman} and Blanchet and Sch\"afer
\cite{BlanchetSchafer}. The
second post-Newtonian correction, at order $v^4$, was
calculated in 1994 by Blanchet, Damour, Iyer, Will,
and Wiseman \cite{BDIWW}. Finally, the spin-orbit and spin-spin
corrections were calculated in 1993 by Kidder, Will,
and Wiseman \cite{KWW}.

We see from Eq.~(\ref{31}) that the post-Newtonian corrections
bring a number of additional source parameters into the picture.
The waveform no longer depends uniquely upon the chirp mass ${\cal M}$;
it now depends upon the masses $m_1$ and $m_2$ separately,
and upon the spin-orbit and spin-spin parameters (which
stay approximately constant as the system evolves toward
coalescence). These new parameters must be included into
$\vec{\mu}$ when the signal is analyzed using matched filtering.

What we have at this point is an expression for the
waveform which is accurate to second post-Newtonian
order. The question facing us is whether this waveform
is sufficiently accurate to be used as measurement
templates. In other words, are the systematic errors
generated by these templates guaranteed to be smaller
than the statistical errors?

Evidently, to answer this question is difficult,
since we do not have access to the exact waveform in
order to make comparisons. In the next section we will
consider a model problem for which the waveform {\it can}
be calculated exactly, thereby enabling us to judge
the accuracy of the post-Newtonian expansion. We will
find, in Sec.~15, that the answer to this question is,
most likely, no: the second post-Newtonian waveform is
not sufficiently accurate.

\section{Waveform calculations: perturbation theory}

A different line of attack for solving Einstein's
equations for a compact binary system is to assume
that one of the bodies is very much less massive
than the other \cite{Poisson}. We therefore demand
\begin{equation}
\mu / M \ll 1,
\label{34}
\end{equation}
where $\mu$ is the reduced mass and $M$ the total mass.
In this limit $\mu$ is practically equal to the smaller
mass $m_1$, and $M$ is practically equal to the larger
mass $m_2$. In contrast with the post-Newtonian approach,
we assume {\it nothing} about the size of the
velocity $v$.

This approach takes advantage of the fact that when
Eq.~(\ref{34}) is valid, the smaller mass creates only
a small perturbation in the gravitational field of the
larger mass, which can be taken to be the
Schwarzschild field. The
total gravitational field can therefore be written as
\begin{equation}
g_{\alpha\beta} = g_{\alpha\beta}^{(0)} + h_{\alpha\beta},
\label{35}
\end{equation}
where $g_{\alpha\beta}^{(0)}$ represents the background
Schwarzschild metric, and $h_{\alpha\beta}$ the perturbation.

Writing Einstein's equations for $g_{\alpha\beta}$ and
linearizing with respect to $h_{\alpha\beta}$, one finds
that the perturbation must satisfy an inhomogeneous wave
equation in the Schwarzschild spacetime. Schematically,
\begin{equation}
\Box^{\alpha\beta\mu\nu} h_{\mu\nu} = 8\pi T^{\alpha\beta},
\label{36}
\end{equation}
where $\Box^{\alpha\beta\mu\nu}$ is an appropriate
curved-spacetime wave operator, and $T^{\alpha\beta}$
the stress-energy tensor associated with the orbiting mass.

We will specifically assume that the central body is
a Schwarzschild black hole of mass $M$. This assumption
is made for simplicity, and removes the need to model
the star's interior. As a matter of fact, the internal
structure of the bodies is irrelevant, except during
the last few orbital cycles before coalescence;
this was discussed in Sec.~5. Taking advantage
of this, we model the orbiting body as a point
particle of mass $\mu$, so that its stress-energy tensor
is a Dirac distribution with support on the particle's world
line. For simplicity, and also because it is physically
well motivated (as explained in Sec.~5), we take the world line
to be a circular geodesic of the Schwarzschild spacetime.
With $\{t,r,\theta,\phi\}$ as the usual Schwarzschild
coordinates, $r_0$ denotes the orbital radius, and
$\Omega = d\phi/dt$ is the angular velocity. We have
\begin{equation}
v \equiv \Omega r_0 = (M/r_0)^{1/2} = (M\Omega)^{1/3},
\label{37}
\end{equation}
and the gravitational-wave frequency $f$ is given by
$2 \pi f= 2\Omega$. We stress once more that in the
perturbation approach, $v$ is not required to be
small. The only limitation on $v$ comes from the
fact that for $r_0 \leq 6M$, circular orbits are
no longer stable; this implies $v < 6^{-1/2}
\simeq 0.4082$.

Black-hole perturbations are conveniently treated with the
Teukolsky formalism \cite{Teukolsky},
in which gravitational perturbations
are represented by the complex-valued function $\Psi_4$,
a particular component of the perturbed Weyl tensor;
the tensor $h_{\alpha\beta}$ can be reconstructed from $\Psi_4$.
The equation satisfied by $\Psi_4$ admits a separation of
the variables. When $\Psi_4$ is expanded in spherical
harmonics and decomposed into Fourier modes $e^{-i\omega t}$,
one obtains an ordinary second-order differential equation ---
the Teukolsky equation --- for the radial function $R_{\omega
\ell m}(r)$. Here, $\ell$ and $m$ are the usual
spherical-harmonic indices. Schematically, and omitting the
subscript $\omega \ell m$, this equation takes the form
\begin{equation}
{\cal D} R(r) = T(r),
\label{38}
\end{equation}
where $\cal D$ is a second-order differential operator, and
$T(r)$ the source, constructed from the particle's
stress-energy tensor \cite{Poisson}.

Equation (\ref{38}) can be integrated in the standard
way by constructing a Green's function $G(r,r')$ out
of two linearly independent solutions to the
homogeneous equation. These solutions, $R_<(r)$ and
$R_>(r)$, respectively satisfy appropriate boundary
conditions at the inner ($r=2M$) and outer
($r=\infty$) boundaries. Schematically, $G(r,r') =
R_<(r_<) R_>(r_>)$, where $r_<$ ($r_>$) is the
lesser (greater) of $r$ and $r'$. The
solution to Eq.~(\ref{38}) can then be expressed
as $R(r) = \int G(r,r') T(r')\, dr'$, and $\Psi_4$
can be reconstructed by summing over all the modes.
Finally, the gravitational waveform $h$ and the
luminosity $dE/dt$ can be obtained from the
asymptotic behaviour of $\Psi_4$ when $r\to\infty$.

To integrate Eq.~(\ref{38}) therefore reduces to
solving the homogeneous Teukolsky equation,
${\cal D} R(r)=0$, for the functions $R_<(r)$
and $R_>(r)$. This, it turns out,
is equivalent to integrating the Regge-Wheeler
equation \cite{ReggeWheeler}
\begin{equation}
\Biggl\{
\frac{d^2}{dr^{*2}} + \omega^2 -
\biggl(1 - \frac{2M}{r} \biggr)
\biggl[ \frac{\ell(\ell+1)}{r^2} - \frac{6M}{r} \biggr]
\Biggr\} X_{\omega\ell}(r) = 0,
\label{39}
\end{equation}
for the functions $X_<(r)$ and $X_>(r)$; here,
$d/dr^* = (1-2M/r) d/dr$. This comes about because
a solution to the homogeneous Teukolsky equation can easily
be related to a solution to the Regge-Wheeler equation. The
relation is known as the Chandrasekhar
transformation \cite{Chandra}.

\section{Luminosity from the perturbation approach}

The problem of calculating the gravitational waveform for
a compact binary system with small mass ratio can
therefore be reduced
to the simple one of integrating Eq.~(\ref{39}). The relevant
dimensionless parameter entering this equation is $M\omega$,
where, for circular orbits, $\omega = m\Omega$
\cite{Poisson}. We therefore
have, using Eq.~(\ref{37}), $M\omega = m v^3$.

For arbitrary values of $M\omega$ the Regge-Wheeler equation must
be integrated numerically \cite{paperII}. This must be
done separately for each selected value of
$v$ in the interval $0.1 < v < 0.4$ (approximately corresponding,
for systems of a few solar masses, to the frequency interval
$10\ \mbox{Hz} < f < 1\ 000\ \mbox{Hz}$). \cite{Reif}

\begin{figure}[t]
\special{hscale=70 vscale=70 hoffset=50.0 voffset=-500.0
         angle=0.0 psfile=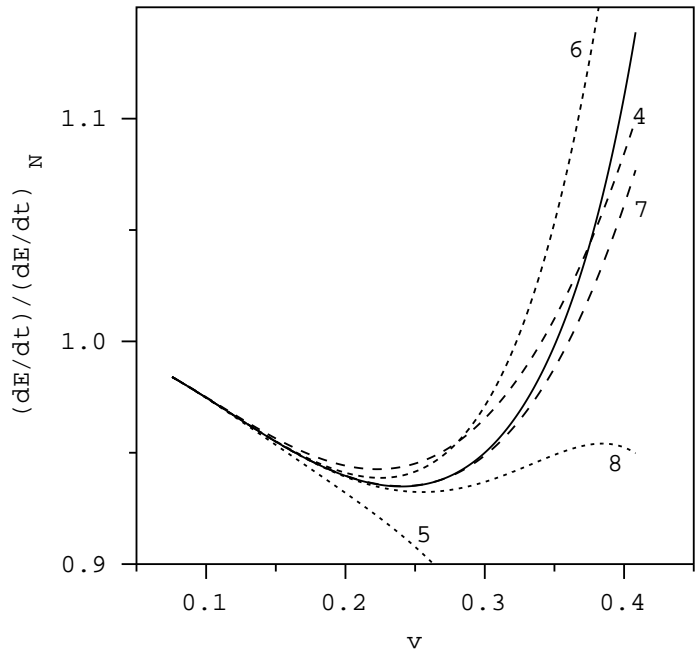}
\vspace*{3.2in}
\caption[Fig. 2]{Various representations of
$(dE/dt)/(dE/dt)_N$ as a function of orbital
velocity $v$. The solid curve represents the exact
results, obtained numerically. The various broken curves
represent the various post-Newtonian approximations, as
explained in the text.}
\end{figure}

Once again we will focus on $dE/dt$, the gravitational-wave
luminosity. In Fig.~2 we present a plot of $(dE/dt)/(dE/dt)_N$
as a function of $v$. This is the ratio of the luminosity as
calculated exactly (numerically) using the perturbation
approach, to the quadrupole-formula expression
\begin{equation}
\biggl( \frac{dE}{dt} \biggl)_{\!\!N} =
- \frac{32}{5} \Bigl(\frac{\mu}{M} \Bigr)^2\, v^{10}.
\label{40}
\end{equation}
The numerical results are depicted as a solid curve.
Apart from negligible numerical errors, these results are
{\it exact} to all orders in $v$; the only assumption used
in the calculation concerns
the smallness of $\mu/M$. The various broken curves will be
described below. The figure shows that $(dE/dt)/(dE/dt)_N$
tends toward unity as $v\to 0$, and stays within
15\% of unity everywhere in the interval $0 < v < 6^{-1/2}$.

It is now easy to judge the accuracy of the post-Newtonian
expansion for $dE/dt$ in the limit $\mu/M \to 0$. One simply
evaluates Eq.~(\ref{31}) in this limit, with $\mbox{\small SO} =
\mbox{\small SS} = 0$, and compare with the numerical results.
This post-Newtonian curve is labeled ``4'' in the figure, and we
see that the numerical results are only imperfectly reproduced.
The consequences of this will be discussed in the next section.

One can in fact do better than this. The perturbation approach
is not only suitable for exact, numerical computations. By
combining it with a slow-motion approximation --- putting
$v \ll 1$ on top of $\mu/M \ll 1$ --- it also becomes suitable
for approximate, analytical computations. Methods for integrating
the Regge-Wheeler equations analytically in the limit
$M\omega \ll 1$ were devised by various authors
\cite{paperI,Poisson,Sasaki}.
The basic idea is to proceed by iterations:
Eq.~(\ref{39}) with $M\omega=0$ is solved
in terms of spherical Bessel functions, and this zeroth
order solution is used as input for the first iteration.

Using these methods, Tagoshi and Sasaki
\cite{TagoshiSasaki} were able to calculate
$dE/dt$ analytically, accurately through fourth post-Newtonian order,
two full orders beyond Eq.~(\ref{31}). They derive the rather
impressive expression
\begin{eqnarray}
\frac{dE}{dt} &=& \biggl( \frac{dE}{dt} \biggr)_{\!\!N}
\Biggl[
1 - \frac{1247}{336}\, v^2 + 4\pi\, v^3
- \frac{44711}{9072}\, v^4
- \frac{8191}{672}\, \pi\, v^5
\nonumber \\ & & \mbox{}
+ \biggl( \frac{6643739519}{69854400} -
          \frac{1712}{105}\, \gamma +
          \frac{16}{3}\, \pi^2 -
          \frac{3424}{105}\, \ln 2 -
          \frac{1712}{105}\, \ln v
   \biggr)\, v^6
- \frac{16285}{504}\, \pi \, v^7
\nonumber \\ & & \mbox{}
+ \biggl( - \frac{323105549467}{3178375200} +
            \frac{232597}{4410}\, \gamma -
            \frac{1369}{126}\, \pi^2 +
            \frac{39931}{294}\, \ln 2 -
            \frac{47385}{1568}\, \ln 3 +
            \frac{232597}{4410}\, \ln v
   \biggr)\, v^8
\nonumber \\ & & \mbox{}
+ \cdots
\Biggr].
\label{41}
\end{eqnarray}
Notice the presence of $\ln v$ terms in this expansion,
as well as that of the Euler number $\gamma \simeq
0.5772$. Notice also that the first four terms reproduce
Eq.~(\ref{31}) in the limit $\mu/M \to 0$, when
$\mbox{\small SO} = \mbox{\small SS} = 0$. This, of
course, is as it should be.

The broken curves in Fig.~2 represent plots of Eq.~(\ref{41})
truncated to various orders in $v$. For example, the
curve labeled ``6'' is a plot of Eq.~(\ref{41}) with all terms
of order $v^7$ and $v^8$ discarded. We see from the figure
that the post-Newtonian expansion converges poorly. (The
suspicion, in fact, is that the series is only asymptotic.)
Witness in particular the poor quality of the curve ``5'' with
respect to ``4'', and compare also ``8'' to ``7''.

\section{Accuracy of the post-Newtonian expansion}

We have seen that in the $\mu/M \to 0$ limit, the post-Newtonian
expansion for $dE/dt$ converges poorly. A similar statement can be
made about $dE/df$: the post-Newtonian expansion
\begin{equation}
\frac{dE}{df} = \biggr( \frac{dE}{df} \biggr)_{\!\!N}
\biggl( 1 - \frac{3}{2}\, v^2 - \frac{81}{8}\, v^4 -
\frac{675}{16}\, v^6 - \frac{19845}{128}\, v^8 + \cdots
\biggr),
\label{42}
\end{equation}
where
\begin{equation}
\biggl( \frac{dE}{df} \biggr)_{\!\!N} =
-\frac{\pi}{3} \mu M v^{-1},
\label{43}
\end{equation}
converges slowly to the exact result \cite{foot2}
\begin{equation}
\frac{dE}{df} = \biggr( \frac{dE}{df} \biggr)_{\!\!N}
\bigl( 1 - 6v^2 \bigr) \bigl( 1 - 3v^2 \bigr)^{-3/2}.
\label{44}
\end{equation}
Contrary to Eq.~(\ref{41}), the expansion (\ref{42})
is actually {\it known} to converge (for all values of
$v$ in the interval $0 < v < 6^{-1/2}$).

In this section we address the issue as to how much of an
obstacle the poor convergence of the post-Newtonian
expansion poses to the construction of accurate
measurement templates \cite{paperVI}.
We will answer this question
for binary systems with small mass ratios, using the
results described in the preceding section. Since there
is no reason to believe that the convergence of the
post-Newtonian expansion would be much improved for
systems of comparable masses, our results should also
apply, at least qualitatively, for such systems.

One way to address this question is to ask, given a
waveform constructed from the exact numerical results,
how much signal-to-noise ratio is lost by matched filtering
the exact signal with an approximate post-Newtonian template?

{}From Sec.~8 we know that filtering with the exact signal
$h(t)$ would give $\mbox{\small SNR}|_{\rm max}$, the largest
possible value of the signal-to-noise ratio. On the other
hand, filtering with a post-Newtonian template $h_n$, where
$n$ denotes the order in $v$ to which the expansion is taken
(for example, $n=4$ represents a waveform accurate to
second post-Newtonian order), gives the smaller value
$\mbox{\small SNR}|_{\rm actual}$. From the results of
Sec.~7 and 8 we obtain
\begin{equation}
{\cal R}_n \equiv
\frac{ \mbox{\small SNR}|_{\rm actual} }{
\mbox{\small SNR}|_{\rm max} } =
\frac{ | \langle h | h_n \rangle |}{
\sqrt{ \langle h | h \rangle
       \langle h_n | h_n \rangle }}.
\label{45}
\end{equation}
The detail of how to compute ${\cal R}_n$ is
presented in Ref.~\cite{paperVI}.

The calculation described here
can only be carried out for binary systems with
small mass ratios, because only for these do we have
access to the exact waveform $h(t)$. Nevertheless, in
the following we shall let $\mu/M$ become large, without
altering our expressions for the exact and post-Newtonian
waveforms. This is done without justification, but reflects
the adopted point of view that the quality of the
post-Newtonian approximation should not be appreciably
affected by the finite-mass corrections.

In doing this extrapolation, some thought must be given
as to the interpretation of the ratio $\mu/M$. In the
limit $\mu/M \to 0$ this can be taken to be both the ratio
of the individual masses or the ratio of reduced mass to
total mass. In the case of comparable masses, some choice
must be made. We note that as $\mu/M$ is allowed to grow large,
expressions (\ref{40}) and (\ref{43}) for the Newtonian quantities
$(dE/dt)_N$ and $(dE/df)_N$ must be replaced by expressions
(\ref{29}), {\it which are formally identical}.
This shows that when extrapolating to the
case of comparable masses, $\mu$ is to be interpreted as the
{\it reduced mass}, and $M$ as the {\it total mass}.

\begin{table}[t]
\begin{center}
\begin{tabular}{ccc}
\hline
\hline
$n$ & PN & EXACT \\
\hline
4 & 0.5796 & 0.4958 \\
5 & 0.4646 & 0.5286 \\
6 & 0.7553 & 0.9454 \\
7 & 0.7651 & 0.9864 \\
8 & 0.7568 & 0.9695 \\
\hline
\hline
\end{tabular}
\end{center}
\caption[Table I]{Reduction in signal-to-noise ratio
incurred when matched filtering with approximate, post-Newtonian
templates. The first column lists the order $n$ of the approximation,
the second column lists ${\cal R}_n$ as calculated using
the post-Newtonian approximation for $dE/df$, and the
third column lists ${\cal R}_n$ as calculated using the
exact expression for $dE/df$.}
\end{table}

We quote the results for ${\cal R}_n$ corresponding to a
system of two neutron stars, each of 1.4 $M_\odot$;
these are displayed in the second column of Table 1. We see
that even at quite a high order in the post-Newtonian
expansion, only three quarters of the
signal-to-noise ratio is reproduced by the post-Newtonian
template. This shows that the apparently small discrepancies
between the exact and post-Newtonian results for $dE/dt$
and $dE/df$ provide a serious obstacle to the construction
of accurate measurement templates.

It is interesting to ask how much of the signal-to-noise
ratio would be recovered if $dE/df$ were kept exact instead
of being expressed as a post-Newtonian expansion. The third
column of the table displays the values of ${\cal R}_n$
calculated in this way. We see that most of the signal-to-noise is
recovered: ${\cal R}_n$ can now be as large as 0.9864 instead
of 0.7651. Why does the exact expression for $dE/df$ give
such better results? It can be established
\cite{paperVI} that this has to do
with the following fact: While the exact expression for $dE/df$
correctly goes to zero at $v = 6^{-1/2}$ (at the innermost circular
orbit), its post-Newtonian analogue fails to do so. [For
example, the right-hand side of Eq.~(\ref{42}), truncated
to order $v^8$, goes to zero at $v\simeq 0.4236 > 6^{-1/2}$,
corresponding to a radius $r_0 \simeq 5.572 M$.] When corrected
for this, the post-Newtonian template gives much better
results.

\section{Conclusion}

We therefore see that the poor convergence of the
post-Newtonian expansion is a serious obstacle to
the construction of accurate measurement templates.
Devising ways to extract the
information contained in gravitational waves produced
during the late inspiral of a compact binary system
poses a great challenge to theoretical physicists. It is
not clear that ``simply'' pushing to higher order in
post-Newtonian theory will be enough to produce
sufficiently accurate measurement templates (such that
the systematic errors will be smaller than the
statistical errors). There may be a need for theorists
to develop alternative ways of dealing with this
problem.

It should be stressed that the convergence problem does
not arise when constructing {\it detection templates}.
Indeed, as was discussed in Sec.~9,
there is no particular need for these templates to be
derived from the equations of general relativity. And since
the detection templates need not involve any parameters of
direct physical significance, the notion of systematic errors does
not apply to them. The only requirement for constructing
detection templates is that they should span the appropriate
``signal space'', and that they should do so the most
efficiently.

The poor convergence of the post-Newtonian expansion is
therefore not an obstacle for {\it detecting} gravitational
waves from coalescing compact binaries. It is only an obstacle
for extracting the information that the waves contain. To
overcome this obstacle will undoubtedly be a theorist's
challenge for years to come. This state of affairs is highly
interesting from a historical point of view: Never
before in the history of gravitational physics has experiment
demanded such a high degree of sophistication on theoretical
calculations.

\section*{Acknowledgments}

This work was supported by the Natural Sciences Foundation
under Grant No.~PHY 92-22902 and the National Aeronautics
and Space Administration under Grant No.~NAGW 3874. This
article was completed while visiting the Theoretical Physics
Institute of the University of Alberta; the author is most
grateful to Werner Israel for his warm hospitality.


\begin{thebibliography}{99}
\bibitem{Thorne87} For broader reviews on gravitational waves,
see K.S. Thorne, in {\it 300 Years of Gravitation}, edited by
S.W. Hawking and W. Israel (Cambridge University Press, Cambridge,
1987); in {\it Proceedings of the Snowmass 95 Summer Study on
Particle and Nuclear Astrophysics and Cosmology}, edited by
E.W. Kolb and R. Peccei (World Scientific, Singapore, 1995).
\bibitem{Taylor} R.A. Hulse and J.H. Taylor, Astrophys. J.
{\bf 324}, 355 (1975); J.H. Taylor, Rev. Mod. Phys. {\bf 66},
711 (1994).
\bibitem{LIGO} A. Abramovici {\it et al.}, Science {\bf 256},
325 (1992).
\bibitem{VIRGO} C. Bradaschia {\it el al.}, Nucl. Instrum. \&
Methods {\bf A289}, 518 (1990).
\bibitem{Reif} See, for example, F. Reif, {\it Fundamentals of
Statistical and Thermal Physics} (McGraw-Hill, New-York, 1965).
\bibitem{Helstrom} C.W. Helstrom, {\it Statistical Theory of
Signal Detection} (Pergamon, Oxford, 1968).
\bibitem{Schutz} B.F. Schutz, Nature (London) {\bf 323}, 310
(1986); Class. Quantum Grav. {\bf 6}, 1761 (1989).
\bibitem{foot1} Other configurations, with narrow-band sensitivity
at a chosen frequency, are possible.
\bibitem{Phinney} E.S. Phinney, Astrophys. J. {\bf 380}, L17 (1991);
R. Narayan, T. Piran, and A. Shemi, {\it ibid.} {\bf 379}, L17 (1991);
A.V. Tutukov and L.R. Yungelson, Mon. Not. Roy. Astron. Soc. {\bf 260},
675 (1993).
\bibitem{Kochanek} C. Kochanek, Astrophys. J. {\bf 398}, 234 (1992);
L. Bildsten and C. Cutler, {\it ibid.} {\bf 400}, 175 (1992).
\bibitem{Damour} For an overview, see T. Damour, in {\it 300 Years of
Gravitation}, edited by S.W. Hawking and W. Israel (Cambridge University
Press, Cambridge, 1987).
\bibitem{MTW} See, for example, C.W. Misner, K.S. Thorne, and J.A.
Wheeler, {\it Gravitation} (Freeman, San Francisco, 1973), Chap. 36.
\bibitem{Wainstein} L.A. Wainstein and V.D. Zubakov, {\it Extraction
of Signals from Noise} (Prentice-Hall, Englewood Cliffs, 1962).
\bibitem{CutlerFlanagan} The following discussion is based on C. Cutler
and E.E. Flanagan, Phys. Rev. D {\bf 49}, 2658 (1994).
\bibitem{Finn} L.S. Finn, Phys. Rev. D {\bf 46}, 5236 (1992).
\bibitem{FinnChernoff} L.S. Finn and D.F. Chernoff, Phys. Rev. D
{\bf 47}, 2198 (1993).
\bibitem{Cutleretal} C. Cutler {\it et al.}, Phys. Rev. Lett.
{\bf 70}, 1984 (1993).
\bibitem{Will} For an overview, see C.M. Will, in {\it Relativistic
Cosmology}, Proceedings of the Eighth Nishinomiya-Yukawa Memorial
Symposium, edited by M. Sasaki (Universal Academy Press, Kyoto,
1994).
\bibitem{BlanchetDamour} The following discussion is based on the
work of Blanchet, Damour, and Iyer, expounded in the following papers:
L. Blanchet and T. Damour, Philos. Trans. R. Soc. London A {\bf 320},
379 (1986); Phys. Rev. D {\bf 37}, 1410 (1988); Ann. Inst. H.
Poincar\'e (Phys. Th\'eorique) {\bf 50}, 377 (1989); L. Blanchet,
Proc. R. Soc. London A {\bf 409}, 383 (1987); T. Damour and B.R. Iyer,
Ann. Inst. H. Poincar\'e (Phys. Th\'eorique) {\bf 54}, 115 (1991).
\bibitem{Thorne80} K.S. Thorne, Rev. Mod. Phys. {\bf 52}, 299 (1980).
\bibitem{KWW} L.E. Kidder, C.M. Will, and A.G. Wiseman, Phys. Rev. D
{\bf 47}, R4183 (1993).
\bibitem{PetersMathews} P.C. Peters and J. Mathews, Phys. Rev. {\bf 131},
435 (1963).
\bibitem{WagonerWill} R.V. Wagoner and C.M. Will, Astrophys. J. {\bf 210},
764 (1976); {\bf 215}, 984 (1977).
\bibitem{paperI} E. Poisson, Phys. Rev. D {\bf 47}, 1497 (1993).
\bibitem{Wiseman} A.G. Wiseman, Phys. Rev. D {\bf 48}, 4757 (1993).
\bibitem{BlanchetSchafer} L. Blanchet and G. Sh\"afer, Class.
Quantum Grav. {\bf 10}, 2699 (1993).
\bibitem{BDIWW} L. Blanchet, T. Damour, B.R. Iyer, C.M. Will,
and A.G. Wiseman, Phys. Rev. Lett. {\bf 74}, 3515 (1995).
\bibitem{Poisson} The following discussion is based on the
work of this author and his collaborators. See E. Poisson
and M. Sasaki, Phys. Rev. D {\bf 51}, 5753 (1995) and
references therein.
\bibitem{Teukolsky} S.A. Teukolsky, Astrophys. J. {\bf 185},
635 (1973).
\bibitem{ReggeWheeler} T. Regge and J.A. Wheeler, Phys. Rev.
{\bf 108}, 1063 (1957).
\bibitem{Chandra} S. Chandrasekhar, Proc. R. Soc. London A
{\bf 343}, 289 (1975).
\bibitem{paperII} C. Cutler, L.S. Finn, E. Poisson, and
G.J. Sussman, Phys. Rev. D {\bf 47}, 1511 (1993).
\bibitem{Sasaki} M. Sasaki, Prog. Theor. Phys. {\bf 92},
17 (1994).
\bibitem{TagoshiSasaki} H. Tagoshi and M. Sasaki, Prog.
Theor. Phys. {\bf 92}, 745 (1994).
\bibitem{foot2} This can be derived simply from the equations
for circular geodesic motion in Schwarzschild.
\bibitem{paperVI} E. Poisson, {\it Gravitational radiation from
a particle in circular orbit around a black hole. VI. Accuracy
of the post-Newtonian expansion}, Phys. Rev. D, in press.
\end{thebibliography}
\end{document}